\documentclass[aps,xpre,longbibliography,showpacs,amsmath,amssymb]{revtex4-1}
\usepackage{amsmath,amssymb,latexsym,epsfig,graphics,epsf}
\usepackage{hyperref}
\usepackage{graphics}
\usepackage{float}
\usepackage{footnote}
\usepackage[utf8]{inputenc}
\usepackage{adjustbox,amsmath,amsfonts}
\usepackage{blindtext}
\usepackage[T1]{fontenc}
\usepackage{lpic}
\usepackage{tabularx,booktabs}
\usepackage[dvipsnames]{xcolor}
\newcommand{\be}{\begin{equation}}
\newcommand{\ee}{\end{equation}}

\newcommand{\Fig}[1]{Figure~\ref{#1}}

\begin{document}
\title{Swap Monte Carlo for diatomic molecules}
	\date{\today}
    \author{Till B{\"o}hmer}\email{till.boehmer@dlr.de}
	\affiliation{\textit{Glass and Time}, IMFUFA, Department of Science and Environment, Roskilde University, P.O. Box 260, DK-4000 Roskilde, Denmark}
    \affiliation{Institute of Frontier Materials on Earth and in Space, German Aerospace Center (DLR), Cologne, Germany}
	\author{Jeppe C. Dyre}\email{dyre@ruc.dk}
    \author{Lorenzo Costigliola}\email{lorenzo.costigliola@gmail.com}
	\affiliation{\textit{Glass and Time}, IMFUFA, Department of Science and Environment, Roskilde University, P.O. Box 260, DK-4000 Roskilde, Denmark}
	
\begin{abstract}
In recent years the Swap Monte Carlo algorithm has led to remarkable progress in equilibrating supercooled model liquids at low temperatures. Applications have so far been limited to systems composed of spherical particles, however, whereas most real-world supercooled liquids are molecular. We here introduce a simple size-polydisperse molecular model that allows for efficient thermal equilibration \textit{in silico} with the Swap Monte Carlo method, resulting in an estimated speedup of $10^3-10^6$ at moderate polydispersity (5-10\%). The model exhibits little difference between size-resolved orientational time-autocorrelation functions. \end{abstract}
\maketitle

\section{Introduction}

When a liquid is cooled fast enough to avoid crystallization, the viscosity increases by typically a factor of $10^{15}$ before the system solidifies at the glass transition \cite{kau48,har76,ang95,deb01,dyr06,ber11,hun12,ric15,mauro,alb22}. The relaxation time increases by a similar factor, and the glass state is arrived at when the equilibration time exceeds the cooling time. Glass-forming liquids continue to attract attention from the physics, chemistry, and material-science communities alike because fundamental scientific problems remain unsolved, e.g.: What controls the extreme slowing down? What causes the ubiquitous deviations from single-exponential relaxation? How to describe the physical aging taking place just below the glass transition? To elucidate such challenging questions it is imperative to have realistic model liquids that can be simulated in the ultraviscous state.

This millennium has witnessed unprecedented advances in glass science that give access to data previously thought to be beyond reach. Vapor deposition has made it possible to produce ultrastable glasses that it would take thousands of years to make by cooling from the melt \cite{edi17}. There have also been tremendous advances in computer simulations, both from hardware improvements including the use of graphics processing units (GPUs) \cite{Berthier2023} and from algorithmic advances. For systems of point particles Swap Monte-Carlo (MC) algorithms now allow for numerically generating extremely viscous equilibrium liquid states and ultrastable glasses \cite{Grigera2001,Brumer2004,Fernandez2006, Fernandez2007,Cavagna2012,Gutierrez2015,Ninarello2017, Berthier2019,Parmar2020, Kuechler2023}. This and similar approaches~\cite{Yamamoto2000,Hukushima1996,shi23} are continuously being improved in a field of rapid development. For instance, ultrastable glasses have been achieved by randomly bonding particle pairs resulting in a mixture of atoms and polydisperse binary molecules \cite{Ozawa2023}, by homogenizing the local virial stress to produce ultrastable glasses in simulations~\cite{leo25}, and in numerical studies of metallic glass formers~\cite{Parmar2020,Parmar2020b}.

The Swap MC method allows for unphysical moves where randomly chosen particles are swapped \cite{gri01}. This trick gained renewed attention when Ninarello \textit{et al.} in 2017 introduced a continuous size-polydisperse system of point particles interacting via a soft repulsive pair potential~\cite{Ninarello2017}. This model proved to be highly stable against crystallization and demixing, making it ideal for equilibrating low-temperature supercooled liquid configurations via Swap MC~\cite{Ninarello2017,Berthier2019}. By alternating standard MC moves with particle swaps, the dynamics is accelerated by up to ten orders of magnitude, allowing one to obtain equilibrium configurations below the experimental glass-transition temperature ~\cite{Ninarello2017}. This model has become standard for exploring the physics of deeply supercooled liquids, and studies of it have enabled advances~\cite{Berthier2023} in the understanding of glass physics in relation to, e.g., dynamic heterogeneity~\cite{Berthier2021,Jung2024}, facilitation~\cite{Scalliet2022,Guiselin2022,Herrero2024}, yielding~\cite{Ozawa2018,Richard2020,Singh2020, Yeh2020,Lamp2022,Ozawa2022}, cooperativity~\cite{Pica2024}, vibrational properties~\cite{Wang2019,Xu2024}, and physical aging~\cite{Herrero2023,Chacko2024,Herrero2023b}.

\begin{figure*}
       \includegraphics[width=1.0\textwidth]{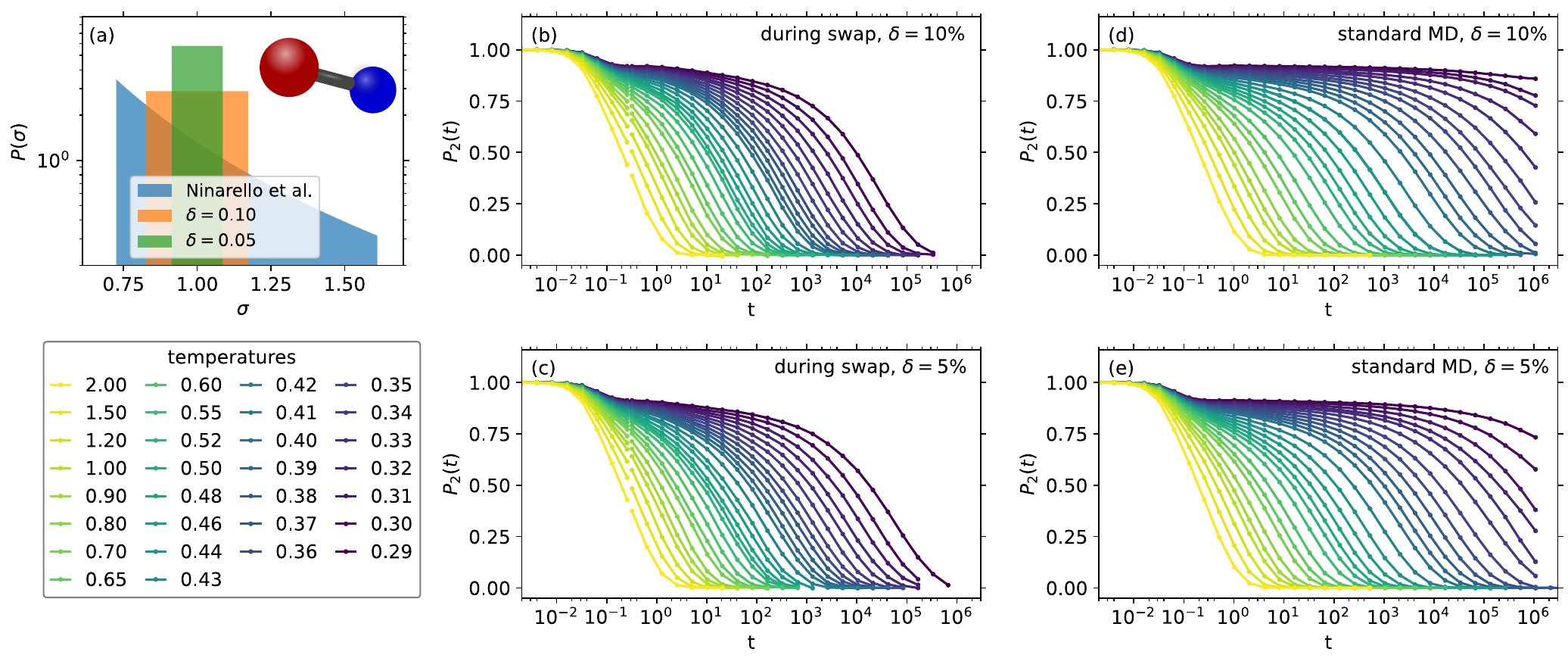}
       \caption{(a) Particle-size distribution of the investigated systems with polydispersities $\delta=5\%$ and $\delta=10\%$. For comparison, we include the distribution for the ``standard'' Swap MC system~\cite{Ninarello2017} with $\delta\approx23\%$ (blue curve). Panels (b)-(e) show orientational time-autocorrelation functions $P_2(t)$ (Eq. (3)) during Swap MC (b,c) and standard MD simulation (d,e) for polydispersity $\delta=10\%$ (upper panels) and $\delta=5\%$ (lower panels) at several temperatures.\label{fig:dynamics}} 
\end{figure*}

\begin{figure}[t]
       \includegraphics[width=0.4\textwidth]{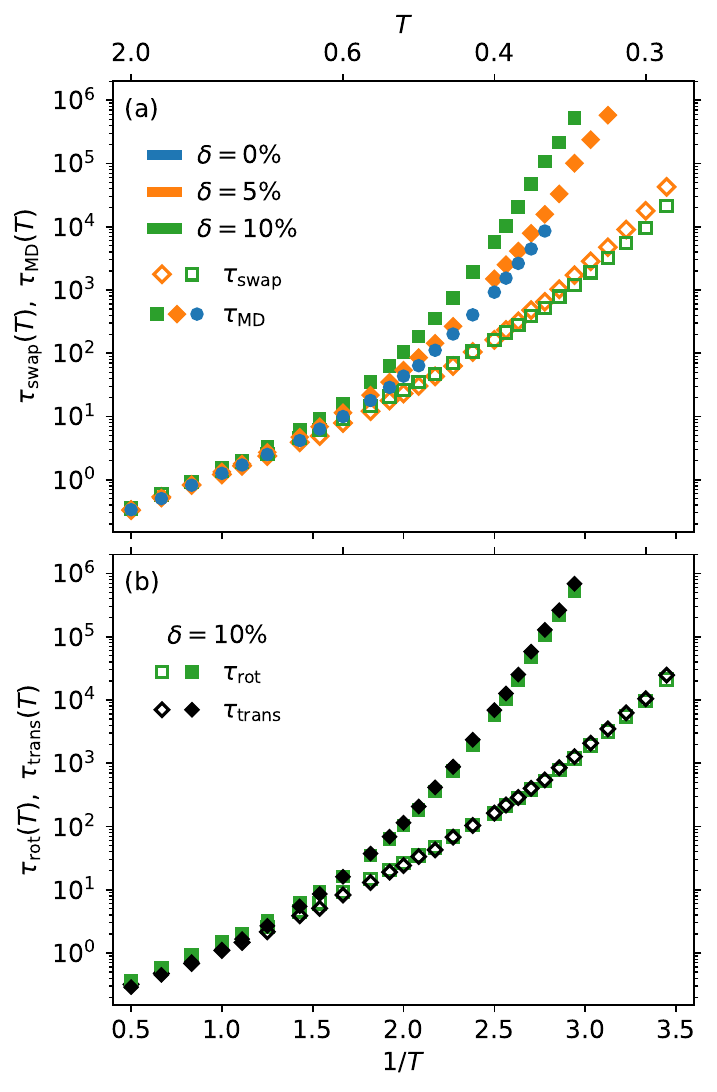}
       \caption{(a) Rotational relaxation times $\tau_\mathrm{Swap}$ (open symbols) and $\tau_\mathrm{MD}$ (full symbols) during Swap and standard MD simulations, respectively, plotted as functions of the inverse temperature. The blue symbols are data for the monodisperse ASD system ($\delta=0\%$) for which Swap of course does not accelerate the dynamics. (b) Comparison of rotational (green symbols) and translational (black symbols) Swap and MD relaxation times for $\delta=10\%$. \label{fig:times}} 
\end{figure}

\begin{figure}
       \includegraphics[width=0.4\textwidth]{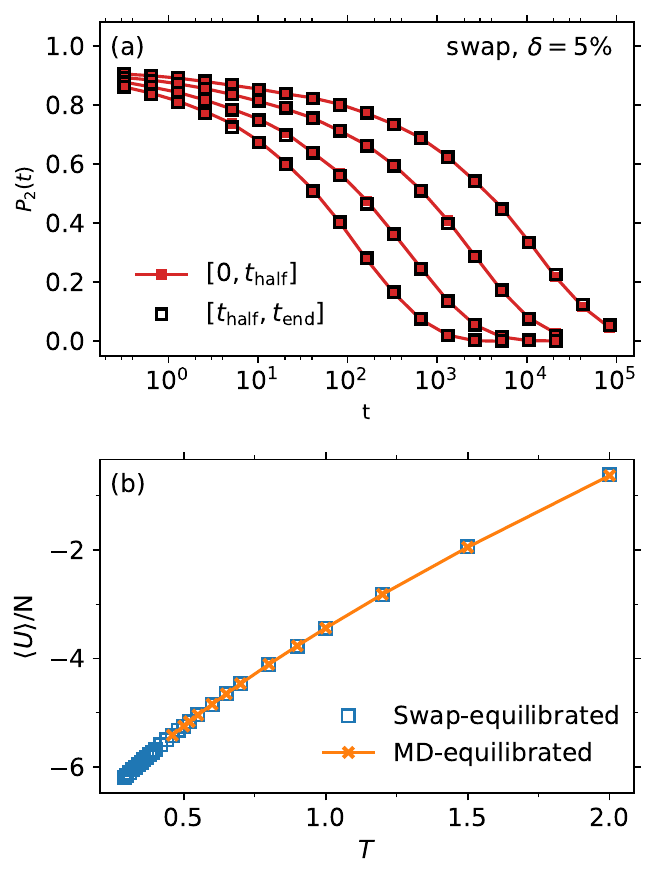}
       \caption{Confirming that the configurations obtained from the equilibration procedure using Swap are equilibrium configurations. (a) illustrates the absence of aging by showing that orientational time-autocorrelation data during Swap averaged over the first half (red symbols) and the second half (black symbols) of the simulation are identical. The figure shows selected data for $T=$0.31, 0.34, 0.38, and 0.42. (b) Average per-particle potential energy $\langle U \rangle/N$ as a function of temperature. The data for Swap-equilibrated configurations (blue symbols) extend the MD-equilibrated data (orange data) to lower temperatures.\label{fig:equ}} 
\end{figure}

\begin{figure*}
       \includegraphics[width=1.0\textwidth]{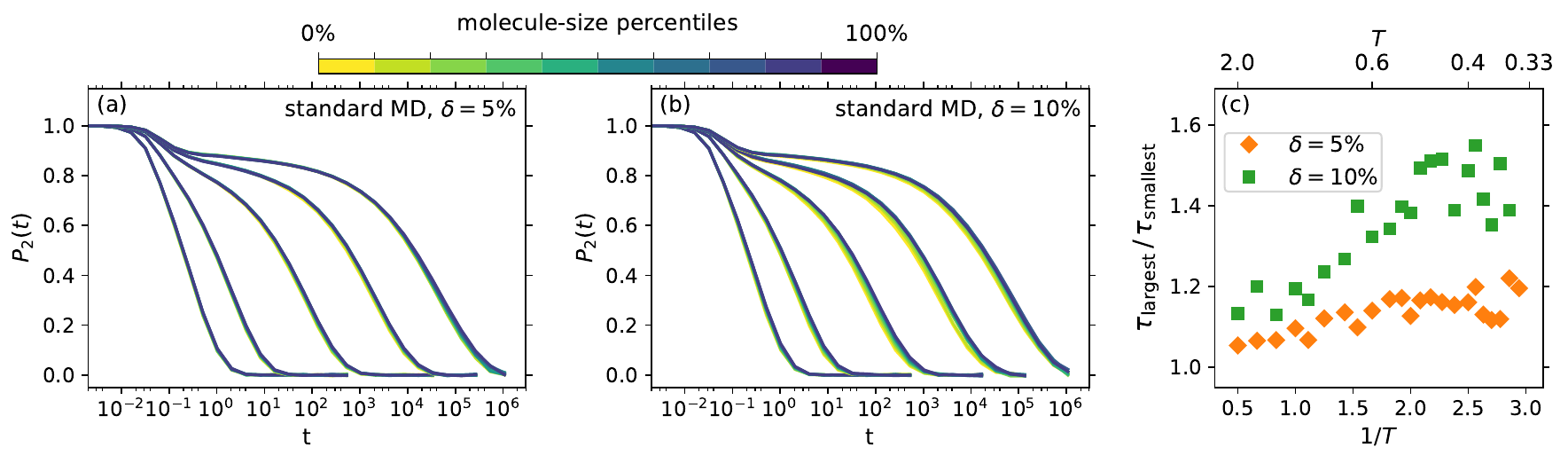}
       \caption{(a) and (b): Size-resolved orientational time-autocorrelation functions from standard MD simulations for $\delta=5\%$ and 10\%. The colors represent data obtained for different 10\%-percentiles of the molecule-size distribution (color bar). From left to right the temperatures are $T=$ 2.0, 0.9, 0.50, 0.40, 0.35 for $\delta=5\%$ and $T=$ 2.0, 0.9, 0.52, 0.42, 0.37 for $\delta=10\%$. (c) The quotient $\tau_\mathrm{smallest}/\tau_\mathrm{largest}$ between the relaxation times of the 10\% smallest and 10\% largest molecules as a function of the temperature.\label{fig:poly}} 
\end{figure*}

Swap MC has so far only been applied to point-particle model liquids. Many real-world supercooled liquids and glasses are composed of molecules, however, and a substantial body of experimental research has documented the distinctive physics of supercooled molecular liquids compared to atomic systems such as metallic glasses~\cite{Lunkenheimer2000,ric15,Kremer2018,boh25}. It is important to be able to model also the \textit{rotational} degrees of freedom that distinguish molecules from point particles. This paper introduces a minimal molecular model, which can be equilibrated efficiently via Swap MC. This allows one to obtain equilibrium configurations near the experimental glass-transition temperature at a reasonable computational cost, thus connecting the world of experimental ultrastable molecular glasses with that of Swap numerical studies.

For a model to capture the physics of real-world supercooled molecular liquids, introduction of size polydispersity would appear not to be realistic. On the other hand, efficient equilibration via Swap MC does rely on introducing polydispersity. That this can lead to significant issues was recently pointed out by Pihlajamaa \textit{et al.}~\cite{Pihlajamaa2023} for the above-mentioned ``standard'' Swap MC system \cite{Ninarello2017} with 23\% polydispersity. At deeply supercooled temperatures, the smallest and largest particles exhibit drastically different behavior; thus their average relaxation times differ by up to a factor of 50 and they, moreover, exhibit markedly different motion mechanisms. While these differences in the size-resolved dynamics can be seen in metallic glasses, they are undesirable for a molecular model. To mitigate such polydispersity-induced effects, we only simulated polydispersities up to $10\%$. As shown below, Swap MC for our molecular model is efficient even for $\delta=5\%$. We show that at such low levels the influence of polydispersity on the dynamics is minimal, demonstrating the viability of our model as a realistic minimal model of deeply supercooled molecular liquids.

\section{Results}

We simulated monodisperse and polydisperse systems composed of asymmetric dumbbell (ASD) molecules at the molecular number density $\rho=0.932$ under periodic boundary conditions. The particle sizes are given by $\sigma_\mathrm{A}=1$ and $\sigma_\mathrm{B}=0.788\sigma_\mathrm{A}$ for the monodisperse system. Each molecule consists of two particles, denoted as A and B (Fig. 1(a)), with masses $m_\mathrm{A}=1$ and $m_\mathrm{B}=0.195$. Particles A and B of the same molecule interact via a harmonic bond of equilibrium length $l=0.584$ and spring constant $k=3000$. The intermolecular interactions are modeled using the Lennard-Jones (LJ) potential, i.e., the potential between particles $i$ and $j$ at distance $r=|\mathbf{r}_i-\mathbf{r}_j|$ is given by
\begin{equation}
    v_{ij}(r) = 4\epsilon_{ij}\left[\left(\frac{r}{\sigma_{ij}}\right)^{-12} - \left(\frac{r}{\sigma_{ij}}\right)^{-6} \right]\,.
\end{equation}
The pair potential is cut at $r=2.5\sigma_{ij}$ using the shifted-force method~\cite{tildesley,tox11a}. 

The size and energy parameters $\sigma_{ij}$ and $\epsilon_{ij}$ follow the standard Lorentz-Berthelot mixing rules, $\sigma_{ij}=(\sigma_i+\sigma_j)/2$ and $\epsilon_{ij}=\sqrt{\epsilon_i \epsilon_j}$, in which $\sigma_i$ and $\epsilon_i$ denote the size and energy parameters of particle $i$. Polydispersity is introduced such that the A and B particle sizes of a given molecule are scaled by the same amount. Thus the B particle is slaved to that of the same-molecule A particle according to $\sigma_\mathrm{B}=0.788\sigma_\mathrm{A}$, while the harmonic bond length is the same for all molecules (this lowers the effective molecular size polydispersity compared to that of the A and B particles). The characteristic energies of the A and B particles are fixed to $\epsilon_\mathrm{A}=1$ and $\epsilon_\mathrm{B}=0.117$. The interaction parameters of the monodisperse system were adopted from earlier works~\cite{ped08a,sch09,ing12b}, where they were designed to provide a minimal model for liquid toluene with the large and small particle mimicking the phenyl and methyl group, respectively.

We study systems with different polydispersity $\delta$ defined by

\begin{equation}
    \delta = \frac{\langle \sigma_A^2 \rangle - \langle \sigma_A \rangle^2}{\langle \sigma_A \rangle^2}
\end{equation}
by choosing $\sigma_A$ to be uniformly distributed as $\sigma_A\in[1-\Delta,\,1+\Delta]$ with $\Delta=\sqrt{12}\delta/2$, which for 5\,\% and 10\,\% polydispersity corresponds to $\Delta=0.087$ and $\Delta=0.173$, respectively. Randomly drawing a value $\sigma_A$ from the distribution for each molecule may lead to substantial finite size effects by sampling the distribution inaccurately~\cite{Kuechler2022}. Therefore, we sampled the distribution inspired by the procedure of K{\"u}chler and Horbach \cite{Kuechler2022,Kuechler2023} by defining 500 different types of molecules, whose $\sigma_A$-values are equally spaced within the interval $[1-\Delta,\,1+\Delta]$. There are four molecules of each size in our system consisting of $N=2000$ molecules.

Two kinds of simulations were performed: 1) Standard \textit{NVT} Molecular Dynamics (MD) simulations with time step $\Delta t=0.002$ and a Nosé–Hoover thermostat with relaxation time $0.2$. The MD units used refer to the size and energy of the A particles of the monodisperse model. 2) Swap MC simulations alternating between short \textit{NVT} MD simulations (for $t=0.32$ each, with $\Delta t=0.005$) and $2N$ consecutive attempts of swapping the parameters $\sigma_\mathrm{A}$ (and, therefore, also $\sigma_\mathrm{B}$) of randomly chosen pairs of molecules. Each swap attempt is accepted with the standard Metropolis-rule probability, resulting in acceptance rates of 15-30\% (SI). To obtain equilibrium configurations, we performed standard MD simulations for at least $500\tau_\mathrm{MD}$ at higher temperatures ($T\geq 0.42$), and Swap MC for at least 50-$200\tau_\mathrm{swap}$ at low temperatures ($50\tau_\mathrm{swap}$ is used for the three lowest studied temperatures). Here, $\tau_\mathrm{MD}$ and $\tau_\mathrm{swap}$ denote the orientational relaxation time during standard MD and Swap. That configurations are truly in equilibrium is confirmed by the absence of aging (see below). We did not observe signs of crystallization in any simulations.

All simulations were performed using \textsc{gamdpy}~\cite{gamdpy}, a Python-only simulation package employing similar optimizations as RUMD~\cite{RUMD} to enable GPU-accelerated MD simulations. Swap moves were implemented on the CPU, however, as they do not benefit from GPU parallelization. Since particle positions remain unchanged during swap attempts, the neighbor list for interaction calculations is reused from the preceding MD sequence, allowing for efficient GPU calculation.

Fig.~\ref{fig:dynamics} illustrates the molecular reorientation dynamics for $\delta=5\%$ and 10\% polydispersity at various temperatures for (b,c) Swap and (d,e) standard MD simulations, quantified via the single-molecule second-order orientational time-autocorrelation function 
\begin{equation}
	P_2(t) = \left\langle 3\cos^2\Theta(t)-1\right\rangle/2\,
\end{equation}
in which $\Theta(t)$ is the angle between the molecule's orientation at time $0$ and time $t$, and $\langle...\rangle$ indicates a moving time-average over all molecules; the time $t$ is the cumulative duration of the short MD sequences in-between the molecule swaps.

For both Swap and standard MD dynamics, the time-autocorrelation functions exhibit a characteristic two-step decay: A low-amplitude short-time decay (relaxation time $\tau_0\approx10^{-1}$) corresponding to the molecules exploring local cages, followed by a long-time decay associated with structural relaxation due to breaking of the local cages. The structural relaxation is significantly accelerated during Swap, especially at low temperatures. The equilibrium data at temperatures $T\lesssim 0.38$ shown in Fig. 1(d,e) could only be obtained by employing Swap for preparing the initial configurations; achieving equivalent equilibration using standard MD would require computational times ranging from months to several years even with GPU-based computations \cite{RUMD}. Equivalent conclusions are drawn from the self-intermediate scattering function quantifying translational dynamics, and from a different orientational autocorrelation function (Appendix).

\Fig{fig:times}(a) summarizes the speedup achieved by Swap relative to standard MD by showing the respective orientational relaxation times $\tau_\mathrm{swap}$ (open symbols) and $\tau_\mathrm{MD}$ (full symbols) as functions of the inverse temperature. Relaxation times are defined via the condition $P_2(\tau)=1/e$. The speedup provided by Swap increases markedly with decreasing temperature. While it obviously decreases with decreasing polydispersity, the speed up remains significant even for $\delta=5\%$.

To estimate the maximum speedup achieved in this study, we examine $\tau_\mathrm{swap}$ and $\tau_\mathrm{MD}$ at the lowest accessible temperature, $T=0.29$. Since $\tau_\mathrm{MD}(T=0.29)$ exceeds the time scale resolvable by standard MD simulations, we extrapolate it using established models for the temperature-dependence of relaxation times. We applied both the Vogel-Fulcher-Tamann (VFT) equation~\cite{Vogel1921} and the parabolic law~\cite{Elmatad2009}, leading to a range of values. For $\delta=10\%$ we find a maximum relative speedup, $\tau_\mathrm{MD}(T=0.29)/\tau_\mathrm{swap}(T=0.29)$, of approximately $10^4-10^6$; for $\delta=5\%$ the range is approximately $10^3-10^4$. Using $\tau_0$ defined above and following the criterion $\tau_\mathrm{MD}(T_\mathrm{g})/\tau_0=10^{12}$ we estimate the experimental glass-transition temperatures to be $T_\mathrm{g}(\delta=10\%)\approx 0.265-0.285$ and $T_\mathrm{g}(\delta=5\%)\approx 0.250-0.270$. This method for estimating $T_\mathrm{g}$ is the same as that used for the above-mentioned standard Swap model \cite{Ninarello2017,Berthier2019}.

In addition to the data for 5\% and 10\% polydispersity, we include in Fig.~\ref{fig:times} also the MD relaxation times of the monodisperse system (blue symbols). There is a mild dependence of $\tau_\mathrm{MD}(T)$ on the polydispersity, i.e., the relaxation time at a given temperature increases slightly with increasing $\delta$. A similar effect was reported by Parmar \textit{et al.} for a modified version of the Kob-Andersen binary LJ mixture~\cite{Parmar2020}. 

\Fig{fig:times}(b) confirms that Swap equally much accelerates the rotational and the translational degrees of freedom, as shown for the $\delta=10\%$ polydispersity model. Here, translational relaxation times $\tau_\mathrm{trans}$ are obtained from the intermediate scattering function $F_\mathrm{s}(q,t)$ evaluated at $q=7.5$, approximately corresponding to the first maximum of the structure factor evaluated for only A particles. 

To confirm that the configurations obtained via Swap are in thermal equilibrium, we demonstrate in Fig.~\ref{fig:equ}(a) the absence of physical aging. For a system that is not in equilibrium, one would observe an explicit dependence of the time-resolved autocorrelation function $C(t_1,t_2)$ on $t_1$, i.e., $C(t_1,t_1+t)$ depends on both $t_1$ and $t$~\cite{Kob1997,boh24}. This behavior is not observed for the Swap-equilibrated configurations; thus $P_2(t)$ averaged over the first half of the simulation, $t_1\in[0,t_{1/2}]$ (red symbols), is indistinguishable from that averaged over the second half, $t_1\in[t_{1/2},t_\mathrm{final}]$ (black symbols). Another confirmation that the systems are in thermal equilibrium involves monitoring the average per-molecule potential energy, which is shown in Fig~\ref{fig:equ}(b) for Swap- (blue) and MD-equilibrated (orange) configurations as a function of temperature. The data for Swap-equilibrated configurations extend the data of the MD-equilibrated configurations to lower temperatures without any noticeable kink or bend as would be observed if Swap did not reach equilibrium ~\cite{Han2020}.

\section{Discussion}

This study has introduced a simple model that can be deeply supercooled and which captures the essential features of molecular liquids like the existence of molecular bonds, anisotropic intermolecular interactions, and rotational degrees of freedom. In terms of the accessible temperature range, Fig.~\ref{fig:times} confirms that equilibrium configurations can be obtained at temperatures comparable to those studied in typical experiments~\cite{Lunkenheimer2000,boh25}, i.e., $\tau_\mathrm{MD}/\tau_0\sim 10^6-10^{12}$ corresponding to real times approaching seconds. 

An important aspect to consider is the role of polydispersity, which is absent in a real molecular liquid. Polydispersity is not expected to significantly alter the underlying physics if the smallest and the largest molecules in the system exhibit the same or very similar dynamics. To test this we study in Figs.~\ref{fig:poly}(a) and (b) size-resolved reorientation dynamics. This is done by plotting $P_2(t)$ for different 10\%-percentiles of the size distribution at selected temperatures. Clearly, the smallest and largest molecules behave similarly. This is confirmed in (c), which shows the quotient of relaxation times of the 10\% largest and 10\% smallest molecules. While there is an increase upon cooling, the ratio eventually saturates and does not exceed 1.2 and 1.5 for 5\% and 10\% polydispersity, respectively. This is in stark contrast to $\tau_\mathrm{largest}/\tau_\mathrm{smallest}\approx 50$ found for point-particle systems with large polydispersity~\cite{Pihlajamaa2023}. We conclude that the ASD system with polydispersity in the range 5-10\% can serve as a minimal model for real-life deeply supercooled molecular liquids.

The presented procedure can be readily extended to more complex molecular models. The efficiency of Swap equilibration is likely to decrease with increasing complexity, however. This is because larger molecules require larger available volume to relax, which can no longer be provided by swapping the sizes of atoms. Nevertheless, the above procedure should be applicable to trimer systems. 

\begin{figure}[h]
       \includegraphics[width=0.45\textwidth]{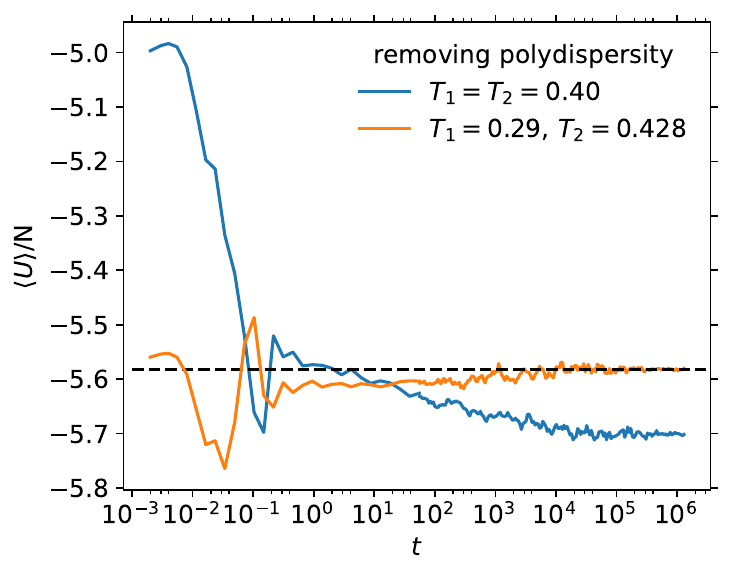}
       \caption{Equilibration following the removal of size-polydispersity for $\delta=5\%$ in two different cases. Details are given in the main text.\label{fig:remove}}
\end{figure}

The use of a system with low polydispersity suggests the possibility of removing the polydispersity without altering the equilibrium. Two attempts in this direction are shown in Fig. ~\ref{fig:remove}. The blue curve shows the relaxation following the removal of polydispersity from a configuration equilibrated at $T=0.40$. After removing polydispersity the potential energy relaxes to a different value of $\langle U \rangle$. The orange line also shows the relaxation of the monodisperse system after the removal of polydispersity, but with the temperature set as discussed below in the next paragraph. In this case, there remains a mild relaxation following the removal of polydispersity, but the system approaches the equilibrium potential energy faster.

A similar observation was recently made on an energy polydisperse system \cite{lan25}. Here the matching temperature is found in a way different from what was previously done, by considering the potential energy as a function of temperature. The potential energy $U_{\delta=0}$ after removing polydispersity can be evaluated from an equilibrated configuration, and by knowing the function $T(U)$ for the monodisperse system it is possible to find the temperature $T(U_{new})$ at which a monodisperse system with potential energy $U_{\delta=0}$ would be in equilibrium. The orange curve in Fig. ~\ref{fig:remove} shows the potential energy relaxation in a simulation with a polydisperse configuration equilibrated at $T=0.29$ as starting configuration. While the positions of particles are taken from the equilibrated configuration, the sizes are all identical (monodisperse) and the thermostat temperature is set to $T=0.428$. The energy between pairs of neighboring particles in a LJ-like system as this one strongly depends on the radii, which may be why removing polydispersity causes a significant difference in equilibrium temperatures. Having a softer system which can be equilibrated via Swap might allow for a removal of polydispersity without a big adjustment in temperature.

In summary, we have introduced a Swap MC procedure for efficiently generating equilibrium supercooled liquid configurations of a simple diatomic molecular model \textit{in silico}, which allows for reaching temperatures approaching that of the experimental glass transition. This was achieved using a system composed of dimers with a minor size polydispersity at the molecule level, in conjunction with an algorithm alternating between swapping the sizes of randomly chosen pairs of molecules and short \textit{NVT} simulations. We explored the acceleration of dynamics in the low-polydispersity limit and found that Swap remains efficient down to 5\% polydispersity. Analyzing the molecule-size-resolved dynamics, we showed that polydispersity in the 5–10\% range only insignificantly alters the qualitative physical behavior of the system. This differs from what is the case for point-particle systems of higher polydispersity~\cite{Pihlajamaa2023}. 

Our findings suggest that the introduced model is a promising candidate for future studies of deeply supercooled molecular liquids, e.g., for comparing to the large body of experimental data on the dielectric relaxation of molecular glass formers \cite{ric15,Kremer2018,boh25}. For instance, it would be interesting to investigate whether there is a $1/\sqrt{\omega}$ high-frequency decay of the $\alpha$-process in the highly viscous phase~\cite{dyr05,dyr06a,nie09,boh25}. Likewise, it would be interesting to monitor whether the polydisperse ASD model upon cooling exhibits decoupling of the rotational and translational degrees of freedom.

\begin{acknowledgments}
The authors are indebted to Mark Ediger for a critical reading of the manuscript and to Thomas Schr{\o}der for providing ideas to optimize the computational efficiency of the Swap algorithm used. This work was supported by the VILLUM Foundation's \textit{Matter} grant (VIL16515).
\end{acknowledgments}

\newpage
\begin{widetext}

\section*{Appendix}

\subsection{Other time-autocorrelation functions}

Figures ~\ref{fig:SI1} and \ref{fig:SI2} show equivalents of Fig.~\ref{fig:dynamics} for the self-intermediate scattering function $F_\mathrm{s}(t)$ and the first-order orientational autocorrelation function $P_1(t)=\left\langle \cos\Theta(t)\right\rangle$, respectively. Both display the same features as found for $P_2(t)$ in the main manuscript; in particular the speedup provided by Swap is identical.

\begin{figure*}[h]
       \includegraphics[width=0.75\textwidth]{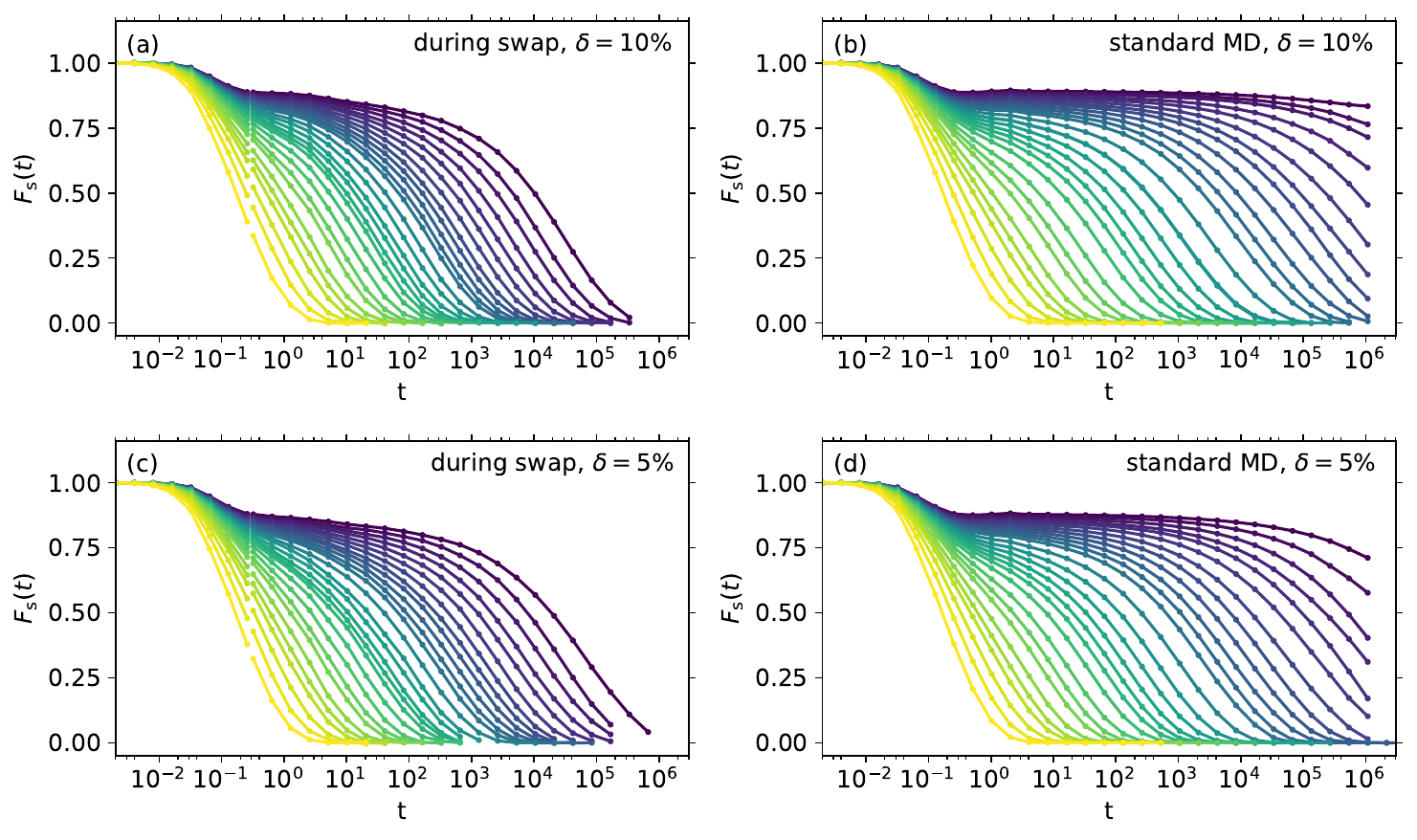}
       \caption{Intermediate scattering function $F_\mathrm{s}(t)$ evaluated for $q=7.5$, approximately corresponding to the first maximum of the $AA$ structure factor, for Swap and standard MD dynamics at different polydispersity. For the temperature color-code see Fig.~1.\label{fig:SI1}} 
\end{figure*}

\begin{figure*}[h]
       \includegraphics[width=0.75\textwidth]{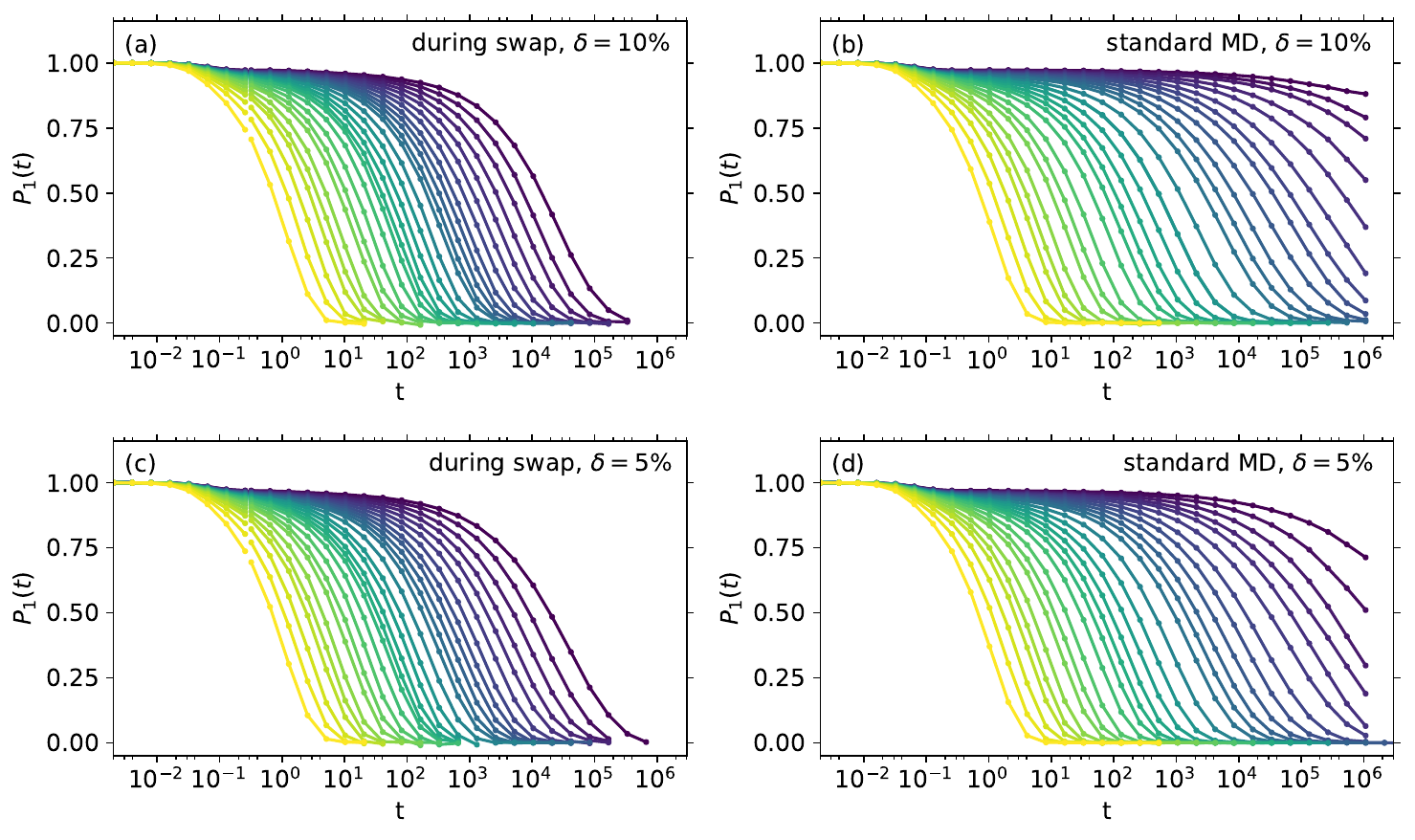}
       \caption{First-order orientational autocorrelation function $P_1(t)=\left\langle \cos\Theta(t)\right\rangle$ for Swap and standard MD dynamics at different polydispersity \label{fig:SI2}} 
\end{figure*}

\subsection{Swap acceptance rate}

\Fig{fig:SI3} displays the average equilibrium acceptance rate for molecule swaps as a function of temperature for both $\delta=5\%$ and 10\% polydispersity. The values are of the same order as the acceptance rates for the standard point-particle Swap model~\cite{Berthier2019,Kuechler2023}.

\begin{figure}[h]
       \includegraphics[width=0.4\textwidth]{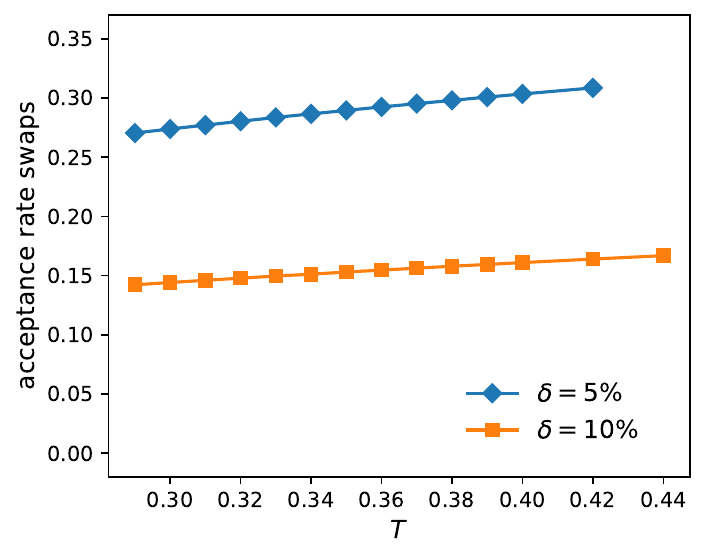}
       \caption{Average equilibrium acceptance rate for molecule swaps as a function of temperature for 5\% and 10\% polydispersity.\label{fig:SI3}} 
\end{figure}


%

\end{widetext}
\end{document}